\documentclass[conference]{IEEEtran}
\IEEEoverridecommandlockouts
\usepackage{cite}
\usepackage{amsmath,amssymb,amsfonts}
\usepackage{algorithmic}
\usepackage{graphicx}
\usepackage{textcomp}
\usepackage{xcolor}
\usepackage{pgfplots}
\usepackage{pgfplotstable}
\usepackage{hyperref}

\def\BibTeX{{\rm B\kern-.05em{\sc i\kern-.025em b}\kern-.08em
    T\kern-.1667em\lower.7ex\hbox{E}\kern-.125emX}}

\makeatother

\begin{document}

\title{Integrating Artificial Open Generative Artificial Intelligence into Software Supply Chain Security\\
{
\footnotesize}
\thanks{}
}

\author{\IEEEauthorblockN{Vasileios Alevizos}
\IEEEauthorblockA{\textit{Karolinska Institutet} \\
Solna, Sweden \\}
\and
\IEEEauthorblockN{George A. Papakostas}
\IEEEauthorblockA{\textit{Democritus University of Thrace} \\
Kavala, Greece \\}
\and
\IEEEauthorblockN{Akebu Simasiku}
\IEEEauthorblockA{\textit{Zambia University} \\
Ndola, Zambia \\}
\and
\IEEEauthorblockN{Dimitra Malliarou}
\IEEEauthorblockA{\textit{IntelliSolutions} \\
Athens, Greece \\}
\and
\IEEEauthorblockN{Antonis Messinis}
\IEEEauthorblockA{\textit{HEDNO SA} \\
Athens, Greece \\}
\and
\IEEEauthorblockN{Sabrina Edralin}
\IEEEauthorblockA{\textit{University of Illinois Urbana-Champaign} \\
Illinois, USA \\}
\and
\IEEEauthorblockN{Clark Xu}
\IEEEauthorblockA{\textit{Mayo Clinic Artificial Intelligence \& Discovery} \\
Minnesota, USA \\}
\and
\IEEEauthorblockN{Zongliang Yue}
\IEEEauthorblockA{\textit{Auburn University Harrison College of Pharmacy} \\
Alabama, USA \\}

}

\maketitle

\begin{abstract}
While new technologies emerge, human errors always looming. Software supply chain is increasingly complex and intertwined, the security of a service has become paramount to ensuring the integrity of products, safeguarding data privacy, and maintaining operational continuity. In this work, we conducted experiments on the promising open Large Language Models (LLMs) into two main software security challenges: source code language errors and deprecated code, with a focus on their potential to replace conventional static and dynamic security scanners that rely on predefined rules and patterns. Our findings suggest that while LLMs present some unexpected results, they also encounter significant limitations, particularly in memory complexity and the management of new and unfamiliar data patterns. Despite these challenges, the proactive application of LLMs, coupled with extensive security databases and continuous updates, holds the potential to fortify Software Supply Chain (SSC) processes against emerging threats.
\end{abstract}


\section{Introduction}

Supply chain security (SCS) is vital because it directly impacts the integrity of products or services, the privacy of data, and the completion of transactions \cite{burmeisters_security_2009}, all of which can have significant operational \cite{asamoah_antecedents_2022}, financial, and reputational consequences \cite{asamoah_antecedents_2022}. When systems are breached, operations can be damaged, disrupted, or destroyed, leading to costs, inefficient delivery, and loss of intellectual property. Human factors, particularly the role of developers, are crucial in securing the Software Supply Chain (SSC) \cite{fourne_viewpoint_2023}. Historically, vulnerabilities in the SSC have been exploited by threat actors who target systems with known weaknesses \cite{last_using_2015}. SSC encompasses both virtual and physical elements. The virtual aspect focuses on protecting information and data within the supply chain, while the physical side deals with the security of goods and products as they transit through the supply chain. These aspects are deeply interconnected, creating a complex web of challenges and considerations.

Each link in the supply chain, from sourcing to disposal, plays a critical role in maintaining security. At the sourcing stage, organizations must identify and select suppliers that align with their quality, ethical, and security standards. During procurement, managing contracts and ensuring timely deliveries is key. In manufacturing, security measures such as access control, data protection, and quality control are imperative. The transportation stage requires strategies to protect shipments from theft, tampering, and cyber-attacks. Distribution involves managing warehouse operations and implementing security measures to protect inventory. In the resale and logistics stage, protecting data and preventing counterfeit products are crucial. Finally, responsible disposal adheres to environmental and data protection regulations. Securing these links involves a comprehensive approach, including risk assessment, due diligence in evaluating suppliers, vulnerability management, and strong access controls \cite{waters_supply_2011}. Data protection, employee training, incident response plans, and regular auditing are essential components of a robust security strategy. 

The importance of SCS lies in its ability to protect against disruptions, ensure quality control, prevent counterfeiting and theft, maintain regulatory compliance, manage risks, and build customer trust. In an increasingly globalized and technologically advanced world, the complexity of supply chains makes their security not just an operational necessity but a critical factor in sustaining a brand's reputation and trust.
As businesses navigate the complexities of global supply networks, artificial intelligence (AI) is becoming a cornerstone in bolstering SCS. This shift towards AI integration reflects a growing confidence in technology's ability to address a spectrum of challenges, from cyber threats to physical interruptions and even issues like counterfeiting. 

Emerging capabilities to sift through enormous data sets, spot trends, and predict future scenarios is increasingly vital in protecting supply chains. It offers an eagle-eye view of goods and information flowing across networks, highlighting unusual activities that might signal a looming threat. For example, automated pipelines can raise red flags over peculiar shipments, spot inconsistencies in inventory, and expose fraudulent schemes. With autonomous system issues are not just stop; it also takes a proactive stance in risk management. By  relying past data, keeping an eye on current happenings, and staying abreast of industry shifts, AI can identify potential vulnerabilities and suggest actions to strengthen the supply chain. Strategies might include broadening the supplier pool, ramping up cybersecurity, or improving emergency plans. This forward-thinking approach arms businesses against emerging dangers and lessens the fallout from disruptions. 

Moreover, emerging technologies significantly enhances decision-making and resource distribution within supply chains. It can pinpoint the most efficient shipping routes, judiciously allocate resources, and maintain optimal inventory levels. This smart, data-driven strategy ensures supply chains run smoothly, reducing costs and boosting overall performance. AI's real-time monitoring and alerting capabilities are particularly crucial in thwarting cybercriminals and other adversaries. It tirelessly watches over network traffic, parses sensor data from facilities, and tracks goods movement, swiftly alerting teams to potential issues. This ability to respond rapidly is key in neutralizing threats swiftly and reducing their impact. Integrating AI into SCS marks a transformative shift in how organizations protect their vital assets. Adopting AI's tools transforms supply chains into resilient and agile operations, well-equipped to adapt to changing threats and disturbances efficiently. As AI's role in SCS continues to expand, it will become an essential element for businesses aiming for top-tier security and operational effectiveness in their supply networks. In literature, there is room for research when it comes to exploring the effectiveness of open source LLMs in identifying security vulnerabilities within software supply chains. Predicting sophisticated patterns could determine how these LLMs can be applied to detect recurring or common flaws in software components.

In software development cycle, we start with the Software Development Life Cycle (SDLC), a vital process for crafting software that's not only effective but also robust and of superior quality. At the outset, we engage with stakeholders, pinpointing their needs through a detailed discussion, a step crucial for laying a strong foundation. Gathering requirements follows, a meticulous process where we dive deeply to understand the precise needs for the software. Moving forward, the design phase springs into action. Here, we explore a variety of design methodologies, including modular and object-oriented designs. Prototyping plays a key role, helping us to confirm that our design choices are on point and align with the requirements. Then comes the heart of the process: software development. In this phase, the focus is on writing clean, maintainable code, a practice that's decisively supported by a range of development tools and Integrated Development Environments (IDEs). As we stitch different software modules together, integration becomes paramount. This is coupled with rigorous testing, including unit, integration, and system testing, ensuring the software runs smoothly. Deployment and maintenance are the next crucial steps. Here, strategies vary, from continuous deployment to phased rollouts, all aimed at ensuring a smooth transition to usability. Maintenance, although sometimes seen as a negligible part, is crucial for the software’s longevity. Moreover, we can't overlook the importance of monitoring and evaluating the software post-deployment. Performance metrics and user feedback keep us abreast of how the software fares in the real world. Additionally, the software development process is never static. Agile methodologies prompt us to continually refine and improve the software, adapting to new needs and technologies. In this journey, each stage of the SDLC is congruent with the next, creating a cohesive and comprehensive process that guides software from conception to completion. This journey is ever evolving, keeping pace with technological advances, and changing user requirements.
In the intricate tapestry of software development, weaving security into every thread of the process is not just essential; it's a categorical imperative. 

From the very start, during the planning and analysis phase, we are on the lookout for potential security hazards. This is not just about ticking boxes for security requirements; it is about a deep, epistemic understanding of what it takes to make a software project not just robust, but impervious to threats. When it comes to design, we are not just building functionality, but fortifying it. This means adopting design principles that are inherently secure and using threat modeling techniques that can forecast and cajole potential risks into the open. As we dive into the coding phase, the mantra is clear: avoid vulnerabilities at all costs. It's a dance of precision and foresight, ensuring that common threats like SQL injection or buffer overflows are relegated to the annals of what not to do. Code review here is not just a cursory glance; it is an empirical, thorough analysis that ensures the code is not just functional but fortified. And when it comes to testing and validation, this is where our security efforts are put through their pace. Think of it as a stress test, using methodologies like penetration testing and the latest automated tools to probe and prod the software, seeking out any chink in its armor. Then comes the deployment and maintenance – a stage often overlooked but crucial. It is not just about deploying software; it is about ensuring it is a secure launch. Patch management here is key; it is like maintaining a diachronic vigilance, constantly updating, and improving to stay ahead of potential threats. And let us not forget the ongoing monitoring and evaluation. This is where we stay alert, eyes peeled for any anomalies or security breaches, ready to respond at a moment's notice. In this journey, embracing a security culture is non-negotiable. It's about ingraining security best practices in the very DNA of the development team, integrating it into the DevOps culture. This holistic approach to security in software development isn't just a strategy; it is a necessity in an ever-evolving security landscape.

This work aims to conduct examination of the capabilities and limitations of freely available LLMs in identifying flaws within the software supply chain. The primary research questions dive into the effectiveness of LLMs in detecting software supply chain vulnerabilities, evaluating the feasibility of replacing traditional static and dynamic scanners, which operate on predefined rules and patterns. Furthermore, it investigates whether this pipeline can accurately handle complex patterns related to software security issues. Through these questions, we seek to contribute analysis into the role of LLMs in enhancing software security. It is hypothesized that employing crowd sourcing generative AI agents, into software SCS can potentially supplant traditional static and dynamic security scanners that rely on predefined rules. However, the imminent conundrum lies in the significant limitations these LLMs face, particularly in memory complexity and the management of new and unfamiliar data patterns. Divvying up tasks between LLMs and conventional methods may be necessary. This paper is structured with the following chapters: the introduction, establishing the importance of SCS, delineating the operational, financial, and reputational risks associated with vulnerabilities in both virtual and physical supply chain elements. Coming up, the background section provides an in-depth exploration of the theoretical foundations and previous studies related to SCS, focusing on the intersection of human factors and technological safeguards. Next, the methodology outlines the experimental framework employed to assess the effectiveness of LLMs in detecting software vulnerabilities, detailing the use of the TruthfulQA benchmark for evaluating model performance. Following this, the discussion section delves into the empirical findings, comparing the efficacy of various LLMs across different programming languages, and analyzing their architectural strengths and limitations. Finally, the Conclusion synthesizes the study's key insights, addressing the limitations encountered and proposing avenues for future research to enhance the resilience of SCS through advanced technologies. This structure ensures a logical progression from theoretical context to empirical analysis, culminating in actionable recommendations.

\subsection{Supply Chain Security (SCS)}

SCS is an intricate domain that intertwines technical and procedural safeguards to protect the integrity, reliability, and trustworthiness of supply chain processes and their components. It emphasizes the importance of establishing trust, utilizing resilient tools, and implementing resilient processes against potential security breaches that can have far-reaching consequences \cite{del_giorgio_solfa_impacts_2022}. In the modern scalable word supply chains are increasingly digital, interconnected, and vulnerable to a myriad of threats. The establishment of trust within the supply chain is paramount, as highlighted by the analysis of software SCS goals which reveal the necessity to verify the identity of participants, the integrity of artifacts, and the reliability of processes \cite{melara_what_2022}. With true each component, whether a piece of software or a physical product, comes from a trusted source and behaves as expected. In the digital world, components that interact with a product have high complexity of verifying their provenance and integrity. Resilient tools are designed or selected with security in mind, capable of identifying vulnerabilities and protecting against attacks. With the emergence of integration Large Language Models (LLMs) to fix bug \cite{clark_enterprise_2002} present an innovative, yet challenging approach to enhancing tool resilience. Resilient processes, on the other hand, focus on the methodologies and workflows that minimize risks and vulnerabilities within the supply chain. This includes automating error-prone manual processes, replacing them with workflows that reduce the attack surface. Many technological providers have invested on supply chain automation solutions to create more resilient supply chain processes \cite{li_large_2023}, \cite{kosasih_review_2023}. Furthermore, the broader perspective on SCS management emphasizes the shift from internal security measures to a comprehensive, end-to-end supply chain protection strategy \cite{burmeisters_security_2009}. Securing every link in the supply chain network is required to measure cross-functionality.

\subsection{Next Generation Software SCS}

With the advent of new threats and vulnerabilities, traditional security measures are no longer sufficient. This calls for a next-generation approach to software SCS that comprehensively understands the chain and accurately identifies where one is positioned within it. Despite the experience in programming languages, security bug could happen inadvertently. Addressing this challenge, recent advancements have explored the use of LLMs models like OpenAI’s Codex and AI21’s Jurassic J-1 for zero-shot vulnerability repair \cite{pearce_examining_2023}. These models, through pre-trained code completion, aim to automate in repairing vulnerabilities without prior explicit examples. However, while promising, this approach faces challenges in coaxing LLMs to generate functionally correct code due to the complexity of natural language and the nuances of coding syntax \cite{pearce_examining_2023}. Furthermore, the application of LLMs extends beyond fixing bugs to revolutionizing cyber threat detection and response. SecurityLLM, a pre-trained model, showcases the utility of LLMs in detecting cybersecurity threats with unprecedented accuracy \cite{ferrag_revolutionizing_2023}. By combining aspects like SecurityBERT for threat detection and FalconLLM for incident response, this model has demonstrated the capability to identify a wide range of attacks with high precision \cite{ferrag_revolutionizing_2023}. Another innovative application is TitanFuzz \cite{deng_large_2023}, which leverages LLMs to enhance the fuzzing of deep-learning libraries. By generating and mutating input programs, TitanFuzz significantly improves code coverage and the detection of bugs. 

In addition to these practical applications, there is ongoing research to better understand and optimize LLMs for software engineering tasks \cite{zheng_towards_2023}. Another demonstration of LLMs to detect vulnerabilities in web applications is KARTAL, addressing the gap in identifying complex, context-dependent security risks \cite{sakaoglu_kartal_nodate}. VulD-Transformer \cite{zhang_vuld-transformer_2023} and VulDetect \cite{omar_vuldetect_2023} models showcase advancements in utilizing deep learning and LLMs for detecting software vulnerabilities. Addressing the challenges in managing complex cloud-native infrastructures, a novel approach using LLMs to analyze declarative deployment code and provide quality assurance recommendations has been proposed \cite{lanciano_analyzing_2023}. Virtual products usually rely partly or entirely on open-source projects allowing more eyes on the security issues, but offering attackers many opportunities. These attacks categorized into different stages from code contribution to package distribution \cite{ladisa_sok_2023}.

\begin{figure}
    \centering
    \includegraphics[width=1\linewidth]{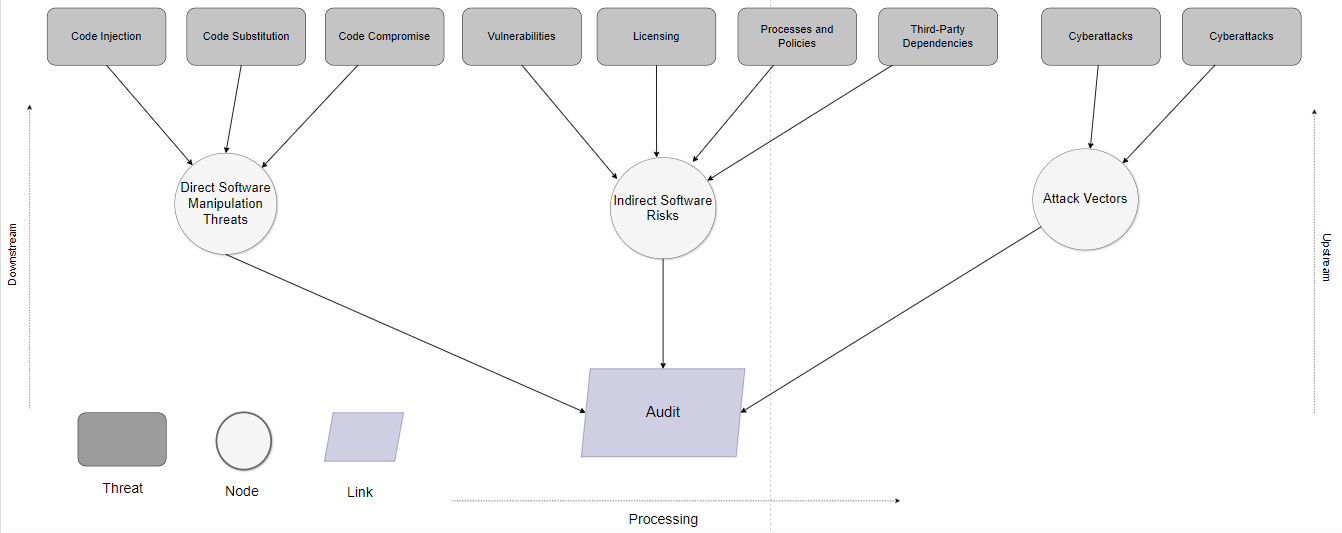}
    \caption{Software Supply Chain Overview of Threats}
    \label{fig:enter-label}
\end{figure}

In the current era, to ensure the safety of modern systems Next Generation Software Supply Chain Security (NGSSCS) facilitates two distinct aspects of securing the software development (see outline Fig. 1). First, capturing trustworthy data is crucial. This includes verifying the identity of entities involved, understanding the components and behaviors of artifacts, and assessing the properties of processes. Additionally, defining clear policies is necessary, which involves establishing a standard format for policies and generating them effectively. Data distribution and verification also play critical roles. It is vital to ensure that data is distributed securely and verified accurately, which includes authenticating trust data and validating it to maintain integrity and security. However, several threats pose risks to the software supply chain. These include code injection, where hackers insert malicious code into software; code substitution, which involves replacing legitimate code with harmful alternatives; and code compromise, which occurs when vulnerabilities or misconfigurations are exploited. Other challenges include vulnerabilities inherent in the software, legal and licensing issues, reliance on third-party dependencies, inadequate security processes and policies, cyberattacks targeting supply chain vendors, and specific attacks like typosquatting and dependency confusion, where attackers trick developers into downloading malicious code by using misleading package names.

\section{Background}

Across multiple studies, the use of LLMs for identifying and reasoning about security vulnerabilities has been explored, showing both promise and limitations. While LLMs can significantly improve vulnerability detection, challenges in model robustness, dataset bias, and generalizability to real-world scenarios remain \cite{ullah_can_2023}. A dataset named DiverseVul \cite{chen_diversevul_2023} is curated by crawling websites for security issues, extracting vulnerable and non-vulnerable source codes from projects. Similar projects try to benefit from existing knowledge bases or combine internet information. But, while increasing training data volume helps, it doesn't necessarily improve the performance in vulnerability detection \cite{chen_diversevul_2023}. In some other study, it was found that transformer-based models, like BERT, were used to accurately classify Java vulnerabilities, achieving up to 99\% accuracy and 94\% f1-score \cite{mamede_exploring_2022}.

\subsection{Taxonomy of Supply Chain Security}

\subsection{Key Factors in Software Supply Chain Safety}

Code quality could be enhanced with third-party libraries, which may have vulnerabilities exploited in supply chain attacks, increasing by 742\% in 2022 \cite{zhang_how_2023}. Commercial LLM services like ChatGPT-4.0 could generate vulnerable dependencies \cite{zhang_how_2023}. Coding automation with LLMs could impact security with risks and threats identified alongside cost-effectiveness \cite{yao_survey_2024}. Conversely, vulnerabilities may be leveraged by malicious actors owing to inherent design deficiencies, the integration of susceptible third-party elements, infiltration and the insertion of malicious code by vendors, undocumented functionalities, alterations in contractual or security premises, shifts in supplier ownership, substandard development practices, insufficient oversight of third-party entities, inadequate data governance, and the provision of excessive access privileges to external parties. Additionally, a plethora of supply chain risks, including reputational, financial, and operational disturbances, further exacerbate these vulnerabilities. The adoption of best practices for SCS is particularly imperative for manufacturing enterprises with intricate supply chains, as they are crucial in mitigating these multifaceted risks.

\section{Methodology}

\subsection{Evaluation}

To assess how well the selected LLMs can detect security issues, including vulnerabilities or outdated code, we apply the TruthfulQA benchmark (as outlined in Figure 2). This method \cite{lin_truthfulqa_2021} checks how truthful and accurate the models are by having them answer questions about a dataset filled with vulnerable or buggy code. The process starts with the model generating answers, which are then measured for accuracy against correct answers and reviewed by humans for truthfulness, particularly for complex or nuanced questions.

\begin{figure}
    \centering
    \includegraphics[width=1\linewidth]{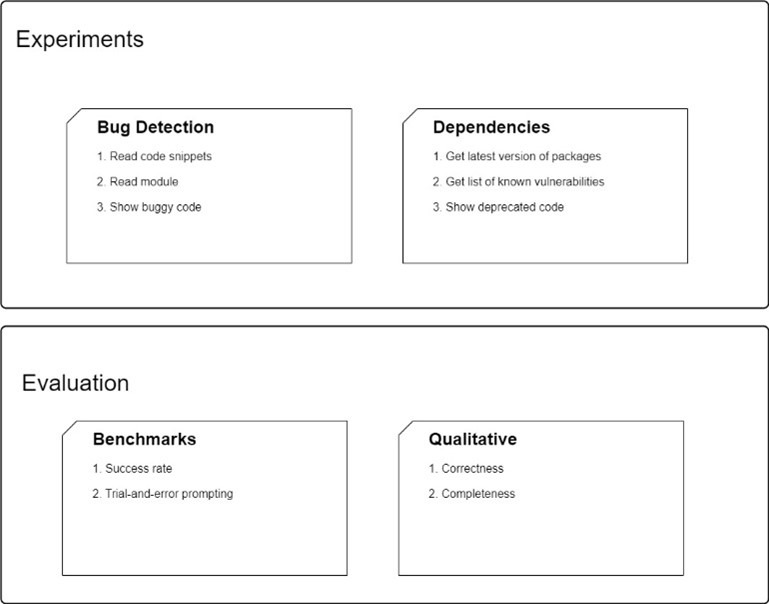}
    \caption{Categorical workflow of our experiments in two different categories of code quality and dependencies. After that, evaluation focuses on the performance of model and overall adaptation to software supply chain. }
    \label{fig:enter-label}
\end{figure}
 
The reliability of the results was evaluated by testing for each prompt through a process of trial-and-error, repeating this testing 10 times for every prompt. The experiments were conducted utilizing a cluster of GPU A100 units. Nonetheless, certain limitations are inherent in the selected LLMs. One significant constraint pertains to the context length, as LLMs are bound by predefined context length parameters. Ensuring that inputs remain compatible with these models necessitates the exclusion of extensive files that surpass these thresholds. The literature also highlights various ethical issues associated with LLM usage, including the provenance of the training data, the substantial energy consumption required for model development, and the concentration of the most advanced models within a limited number of major technology corporations ~\cite{zhang_vuld-transformer_2023}.

\subsection{Dataset}

For the purpose of our accumulation to prove the null hypothesis that the LLMs cannot be used to facilitate software supply chain national vulnerability database \cite{byers_national_2022}. In the selected dataset, an equitable division was performed into almost equal subsets per programming language, each containing 500 vulnerabilities or deprecated anti-patterns. Apropos of ensuring balanced representation, this bifurcation facilitated a comprehensive analysis across different languages.

\section{Discussion}

Detection of vulnerabilities or deprecated code across different programming languages reveal notable differences in the performance and architectural strengths of the models tested. From the outset, OpenLLaMA \cite{openlm2023openllama, together2023redpajama, touvron2023llama} indissolubly stands out with the highest overall performance, particularly excelling in C and Objective-C. One performance factor attributed by the strategy of training data mixture, along with transformer architecture, which allows it to handle tasks from nose to tail with remarkable coherence. In contrast, Gemma \cite{gemma_team_gemma_2024, gemma_2024}, with its strong generalist capabilities and optimized architecture, perpetuates its performance across most languages. With highest score of detection, two technologies stood out. C++ and Python, with an imperceptible demeanor of ease. However, Mistral-7B \cite{jiang_mistral_2023}, despite its balanced performance, remains somewhat inconspicuous in specific languages like Ruby and Perl. This is likely due to its smaller parameter size and limited knowledge retention, which scotches its ability to compete with larger models in retaining extensive language-specific information. Meanwhile, GPT-2 \cite{radford2019language} maintains a demeanor of consistency across various languages, particularly in Python and C++. Yet, its performance in JavaScript and Ruby reveals indescribably subtle limitations in handling more modern or dynamic languages. Finally, Phi2 \cite{abdin2024phi} demonstrated lower scores across most languages, a reflection of its smaller model size and limited factual knowledge storage, thereby scotching its chances to excel in tasks that require extensive and precise language-specific knowledge. The architectural elements, such as attention mechanisms and context length, are indissolubly linked to the models' ability to detect vulnerabilities and deprecated code. Larger models, by their very conception, generally outperform smaller ones due to their capacity to store and process vast amounts of information, making the differences in their capabilities acutely noticeable, though sometimes imperceptible at first glance.

\begin{table}[htbp]
\centering
\caption{Summary of average performance per model with truthfulQA (TQA) of tasks and average detection per task.}
\begin{tabular}{|l|c|c|}
\hline
\textbf{Model} & \textbf{TQA} & \textbf{Avg} \\
\hline
GPT-2 \cite{radford2019language} & 59.34 & 67.02 \\
Gemma, 7B \cite{gemma_team_gemma_2024, gemma_2024} & 52.34 & 66.63 \\
Mistral-7B \cite{jiang_mistral_2023} & 52.39 & 63.93 \\
OpenLLaMA \cite{openlm2023openllama, together2023redpajama, touvron2023llama} & 59.2 & 66.61 \\
Phi2 \cite{abdin2024phi} & 54.34 & 60.38 \\
\hline
\end{tabular}
\end{table}

Traditional matching methods, which are ubiquitous in the industry, rely heavily on predefined rules and specific patterns, often leading to biases based on the datasets they are trained upon. This fastidious adherence to static rules can obfuscate the detection of novel threats, as the inexorable evolution of cyber risks demands more adaptive approaches. Muti-task agents, by incorporating a prismatic spectrum of knowledge, could offer an inclination towards preventing ulnerabilities beyond predefined patterns. At the same time, the practical advantages of using LLMs are limited by their inability to access up-to-date online information, thus leaving them caught between a rock and a hard place when confronting furtive attack patterns. While LLMs represent the quintessence of advanced AI capabilities, without real-time data integration, their potential remains constrained. Moreover, the traditional way handling detections is cognizant of regulatory compliance requirements and provide deterministic results.

\section{Conclusion}

\subsection{Limitations and Future Directions}

It is imperative to comprehend the inherent capabilities and limitations of the model under examination, including its intended functionalities, the domain of its training data, and any documented constraints. Such an understanding is instrumental in crafting testing prompts that are both rigorously challenging and contextually relevant, thereby ensuring the validity of the evaluation process. A notable limitation within our study pertains to the intricate nature of memory as it pertains to the tuning of new information and the integration of unfamiliar patterns within language models. Additionally, an overreliance on prompts devoid of pipeline constraints can precipitate vulnerabilities wherein malicious instructions are covertly embedded, thereby exposing the system to a spectrum of security threats. Nevertheless, these challenges underscore the potential for further exploration. Initiating with language models from their inception or recalibrating existing ones, by leveraging extensive databases replete with security vulnerabilities and safe patterns, presents a promising avenue. This approach, coupled with regular updates reflecting the latest security protocols, could significantly bolster the resilience of such models. Human errors on digital world necessitates a profound degree of critical thinking, with accountability being of paramount importance. At the current stage, LLMs have the potential to function as a valuable augmentation to existing monitoring services, offering an additional layer of coherent and informed recommendations.



\end{document}